\documentclass{ws-procs10x7}

\begin{document}

\title{Why pentaquarks are seen in some experiments and not in others}

\author{Harry J. Lipkin}

\address{Department of Particle Physics,
Weizmann Institute of Science, \\ 
Rehovot, Israel \\ 
E-mail: harry.lipkin@weizmann.ac.il\\
and\\
School of Physics and Astronomy \\
Raymond and Beverly Sackler Faculty of Exact Sciences \\
Tel Aviv University, Tel Aviv, Israel\\
and\\
High Energy Physics Division, Argonne National Laboratory\\
Argonne, IL 60439-4815, USA}
\twocolumn[\maketitle\abstract{
 The $\Theta^+$ Pentaquark is a Very narrow $\Gamma \approx 1$ MeV 
 KN Resonance. Why do some experiments see it and others do not?
The lowest quark configuration
that can describe it is 
 exotic $uudd \bar s$.  
 Why have no exotics been seen before?
 Is this the beginning of a new spectroscopy?
Can it help toteach us about 
How QCD makes hadrons from quarks and gluons?
}]

\def\bra#1{\left\langle #1\right|}
\def\ket#1{\left| #1\right\rangle}
%The allowed length is 4 Pages                                                                                

\section{Introduction} 

\subsection{ A wrong question}
 
People keep asking whether the pentaquark really exists. This is the
wrong question. They recall the a2 splitting and the zeta particle
which went away. But I am now haunted by a different memory, In the
summer of 1964 Jim Smith told me about evidence from their kaon
experiment at Brookhaven for a $2\pi$ decay of the long-lived kaon..
He asked me whether this could be explained without CP violation. I
didn't see how and stupidly believed that there must be something
wrong with their data. But their data  were right and appeared as the
second published evidence for CP violation\cite{jimsmith}. 

      My response to the wrong question about pentaquark existence
is  what I should have told Jim Smith in 1964, Nature is providing us
with new data that theorists do not understand. The right question is
how can we understand the meaning of these data and how can we point
experimentalists in the right direction to clarify the physics which
may be very interesting,

\subsection{Guidance from the wisdom of Feynman and Wigner}
\begin{enumerate}
 \item Feynman told us that we learned from sharpening contradictions.
\item Wigner told us that a few free parameters can fit an elephant. 
A few more can make him wiggle his trunk
 \item Wigner's  response to questions about a particular theory he did not
like was: ``I think that this theory is wrong. But  the old Bohr - Sommerfeld
quantum theory was also wrong. Could we have reached the right theory  without
going through that stage? 
\end{enumerate} 

\noindent Apply this wisdom to the pentaquark 
\begin{enumerate} 
 \item Contradictions between different pentaquark experiments 
 are not sharp enough.
\item Too many theoretical calculations blur contradictions with
free parameters.
 \item The diquark-triquark (wrong?) model\cite{NewPenta}
 without many free parameters may lead us in the right direction. . 
\end{enumerate} 
\section{Experimental contradictions between searches for
exotics}

Some experiments see the pentaquark;
 others definitely do not\cite{cryptopen}
. No theoretical model addresses this problem. Comprehensive
review\cite{jenmalt} analyzes different models.

\subsection{Different initial states with different final state overlaps}

\noindent {\bf 1. The new $\bar D_s^* \rightarrow \bar D_s \eta$}  

Seen at SELEX.but not by others .. 

\noindent  SELEX has $\Sigma^-(sdd)$ beam 

$\Sigma^-(sdd) + c \bar c \rightarrow (sd \bar c)_{triquark}$  ...

 $(sd \bar c)_{triquark} + \bar d 
\rightarrow (s \bar c d \bar d)
\rightarrow  \bar D_s \eta $.

\noindent Is $sd$ in $\Sigma^-$ beam needed to produce  $D_s \eta $.?

Evidence that the $s$ quark in the beam goes into the final state is given by 
the charge asymmetry.  $\bar D_s \eta $ is seen; but $D_s \eta $ not.

  $sd$ in   beam can produce  $(s \bar c d \bar d) 
\rightarrow  \bar D_s \eta $,

 \noindent But cannot produce
 $(c \bar s d \bar d \rightarrow D_s \eta)$

\noindent {\bf 2. The $\Theta^+$ pentaquark}  

In low-energy photoproduction experiments the baryon and $\bar s$ antiquark in
the $\Theta^+$  are initially present  in the target baryon and the $s \bar s$
in the photon. In other experiments that do not see the $\Theta^+$. the cost
of baryon-antibaryon and/or $s \bar s$ production by gluons must be used to
normalize the production cross section for comparison with the photoproduction
cross sections

$\gamma p$  sees $\Theta^+$;  $e^+ e^-$ doesn't.
Can the cost of producing $B \bar B$ and $s \bar s$
explain the difference between  $\gamma p$ and $e^+ e^-$?

\subsection{$\Theta^+$ production via $N^*(2400)$ }

A specific production mechanism that may be
present in experiments that see the $\Theta^+$ and absent in
those that do not\cite{cryptopen}.is suggested by CLAS data  on 
$\gamma p \rightarrow \pi^+ K^- K^+ n$, The $(K^+K^-n)$ mass
distribution 
show a peak at the mass of 2.4 GeV that might
indicate a cryptoexotic
$N^*$ resonance with hidden strangeness. Searches for such baryon
resonances\cite{Landsberg:1999wn} have
indicated possible candidates but did not go up to 2.4
GeV. Further $N^*$ resonance evidence is hinted in 
preliminary results\cite{cryptopen} from NA49. 
 
Some experimental checks of this mechanism\cite{cryptopen} are:

\begin{enumerate}
 \item
Experiments which see the $\Theta^+$ should examine the mass spectrum of 
the $K^-\Theta^+$ and $K_s \Theta^+$ systems.
\item 
Experiments searching for the $\Theta^+$ should check for possible
 production of a 2.4
GeV $K^-\Theta^+$ or $K_s \Theta^+$ resonance 
\item
In the photoproduction reactions
 
\noindent $\gamma p \rightarrow N^*(2400)\rightarrow \bar
K^o \Theta^+$ and 

\noindent $\gamma p \rightarrow N^*(2400)\rightarrow \bar
K^{*o} \Theta^+\rightarrow \pi^-
K^+ \Theta^+ $

\noindent the $K$ or $K^*$angular distribution should show no
forward-backward asymmetry. 
\item
If the photoproduction reaction 

\noindent $\gamma p \rightarrow \pi^+ K^- K^+ n$
\cite{cryptopen}
 goes via  $N^*(2400)$.the pion goes forward and
everything else is in the target fragmentation region.

\item Search for other $N^*(2400)$ decay modes. SU(3) predicts  $N^*(2400
\rightarrow \pi^- N^{*^+}$  where $N^{*^+}$ is the nonstrange SU(3) partner of
the $\Theta^+$.  Decays into $K \Lambda$, $K \Sigma$, $K\Sigma^*$ and $\phi N$.
are allowed in some models but auppressed in a diquark-triquark
model\cite{NewPenta}  by the centrifugal barrier against passage of a quark in
the triquark from joining the diquark. .

\item  Both charge states $N^{*+}(2400)$ and $N^{*o}(2400)$  should be observed.

\end{enumerate}
\subsection{Production via meson and baryon exchanges}
In  $\gamma p \rightarrow
 \bar
K^{*o} \Theta^+\rightarrow \pi^-
K^+ \Theta^+ $ and $\gamma p\rightarrow \bar K^o \Theta^+$ 
meson exchange.predicts a forward-peaked $K$ or $K^*$ angular distribution.
Baryon exchange predicts  backward peaking\cite{Karliner:2004qw}
and the same baryon exchange with equal  production 
in \ $\gamma n \rightarrow K^-  \Theta^+$..

$\Theta^+$ production by baryon exchange is related to reactions between
nonexotic hadrons via $\Theta^+$ or other  exotic baryon exchange. An 
appreciable contribution of the
diagram proposed  for $\Theta^+$ photoproduction with
an outgoing backward kaon indicates an appreciable $K N
\Theta^+$ vertex  that should also contribute appreciably to backward $K^-
p$ charge-exchange\cite{Karliner:2004qw}. Some previously ignored backward $K^- p$ charge-exchange
may still be available.

Models\cite{Karliner:2004qw}  which explain the narrow width of the
$\Theta^+$ by a suppressed $N K \Theta^+$ coupling relative to $N K^*
\Theta^+$  predict that $\Theta^+$ production with a backward $K^*$ should be
stronger than the production with a backward kaon. 

\section{QCD Guide to exotic search}
\subsection {BJ's question in 1986}
 
In $e^+ e^- $ annihilation a created $q \bar q$ fragments into
hadrons. 
The $q$ can pick up a $\bar q$ to make a meson 
or a $q$ to make a diquark $(qq)$. The $qq$
can pick up another $q$ to make a baryon 
but might pickup a $\bar q$ to make a ``triquark" ($qq\bar q)$ bound
in a color triplet state. Picking up two more quarks makes a pentaquark

BJ asked: ``Should such states be bound or live long to be observable as
hadron resonances? What does the quark model say?'

The quark model says that a petaquark is a five-body problem with no feasible 
exact solution  The simplest approach  assuming space factorization and
symmetrization explains absence of  low-lying exotics and gives negative parity
pentaquarks, but does not predict $\Theta^+$.
A positive parity $\Theta^+$ suggested  by the the chiral soliton
model\cite{NewPenta} requires a p-wave pentaquark and may imply a stable lower
s-wave  baryon\footnote{If the splitting between  the lowest positive
strangeness s and p-wave baryons is similar to the 290 MeV $\Lambda (1115)$ -
$\Lambda(1405)$  splitting  the s-wave  $\Theta^+$ is a bound KN state well
below the KN threshold. Experimental evidence for and against should be
checked.}. A five-body system has ten possible pairs for p-wave  giving too
many possible states

\subsection{Color-magnetic interaction and flavor antisymmetry}

QCD motivated models\cite{Nambu,LipkTriEx,DGG} show that breakup of exotic
multiquark color  singlet states into two separated color singlets loses no
color electric energy and gains kinetic  energy.

\noindent \begin{equation} 
\ket{singlet} \rightarrow \ket{singlet}  + 
\ket{singlet} \label{singlet} 
\end{equation}

Extending to multiquark states\cite{Jaffe} the remarkably successful\cite{NewPenta} DGG
model\cite{DGG} showed that only the short-range color-magnetic interaction can
produce binding in single cluster or bag models.

The Pauli principle requires flavor-symmetric quark pairs  to be antisymmetric
in color and spin at short distances and therefore to have a repulsive
short-range  color-magnetic interaction. The  best candidates for multiquark
binding should have a minimum number of same-flavor pairs

The nucleon has only one same-flavor pair;  
$\Delta^{++} (uuu)$ has three. Two extra same-flavor pairs 
cost 300 Mev.
Flavor antisymmetry principle\cite{liptet} explained absence of lowlying 
exotics and suggested search for $H$ dibaryon $uuddss$. 
Extension to heavy quarks\cite{liptet} suggested exotic tetraquarks and
anticharmed strange pentaquark\cite{Pent97}  $(\bar cuuds)$ 

Quark model calculations with flavor antisymmetry 
told experimenters to look for  $(\bar c uuds )$ pentaquark with only one 
same-flavor pair; not the 
$\Theta^+(\bar s uud )$.with two.
 Ashery's E791 search for $\bar c uuds$ found events\cite{E791Col}; not 
convincing enough. 
Better searches with good vertex detectors and particle ID\cite{Pent97}.
are needed; One gold-plated event showing a proton emitted from a secondary
vertex and defining a new unknown baryon is enough without
statistics.

Finding the $\Theta^+$ suggests a two-cluster model\cite{NewPenta}. A
diquark-triquark in  p-wave $\ket{ud; ud \bar s}$ with a centrifugal barrier 
separates repulsive identical $uu$ and $dd$ pairs. The $ud$ pairs are bound
fermion pairs, not bosons, with wave functions not uniquely defined and
depending on external environment. The diquark is in an external color triplet
field like the $ud$ pair in the $\Lambda$ and is assumed to have the same
unique wave function and mass. The $ud$ pair interacting with the $\bar s$ has
a very different environment. Two different color-spin couplings and different
$ud$ wave functions arise for the triquark with roughly the same color-magnetic
energy\cite{jenmalt}. These two $ud$ pairs are thus very different anc cannot
be considered as identical bosons\footnote{Treating diquarks whose wave
functions depend upon external environment as elementary  bosons misses
essential physics just like  treating Cooper pairs as bosons misses
supercondctivity. Boson condensates cannot superconduct because any moving boson 
can lose energy by collision with an impurity.  
A moving Cooper pair  in a BCS superconductor cannot lose
energy by collisions. Changing one pair wave function changes
the environment of all other pairs and creates the energy gap which is
the key to BCS superconductivity.}. 

The two nearly degenerate states  $\Theta_1$ and
$\Theta_2$  both coupled to a KN final state are mixed by the loop
diagram\cite{Karliner:2004qw}   $ \Theta_1 \rightarrow KN \rightarrow \Theta_2
$. Diagonalizing the loop diagram gives aproximate mass eigenstates $\Theta_L$
not coupled to $KN$; $\Theta_S$ broad and lost in
continuum\cite{Karliner:2004qw}. Exact calculation of the narrow width depends
upon unknown parameters

The color-mag<netic force keeps the triquark $(ud \bar s)$ stable against
$ud \bar s \rightarrow d + K^+$ breakup, while $\ket{ud; ~ud \bar s}
\rightarrow \ket{udd} + K^+$  crosses centrifugal barrier. A d-wave
$\ket{us; ud \bar s}$ model for $N^*(2400)$  can explain $N^*(2400)
\rightarrow K^o + \Theta^+$  via diquark transition $\ket{us} 
\rightarrow  K^o + \ket{ud}$  

\subsection {Tests for 
 antidecuplet purity} 

Does the antidecuplet  mix with aoctet? 

Experimental tests for a pure $\bar {10}$ $N^*$

$\gamma p \rightarrow N^{*+}$ forbidden; 
$\gamma n \rightarrow N^{*0}$ allowed

\noindent  $ \sigma (\gamma p \rightarrow  K^+ \pi^-\Sigma^{*})=$
  $\sigma (\gamma p \rightarrow K^+ K^- N^{*})=$ 

\noindent  $\sigma (\gamma p \rightarrow \pi^+ \pi^- N^{*+})=$
${1\over 3}\sigma (\gamma p \rightarrow \pi^+ K^- \Theta^+)$ 

  But if $\gamma p \rightarrow \pi^+ N^{*o}(2400) 
 \rightarrow \pi^+ M^- B^+$

\noindent SU(3) breaking gives a factor 3.difference

\noindent $\sigma (\gamma p \rightarrow \pi^+ \pi^- N^{*+})=$ 
$\sigma (\gamma p \rightarrow \pi^+ K^- \Theta^+)$

.

\section*{Acknowledgments}
The original work reported in this talk was in collaboration with Marek
Karliner.
This work was partially
supported by the U.S. Department
of Energy, Division of High Energy Physics, Contract W-31-109-ENG-38
%\bibitem{ja} M. Barranco and J. R. Buchler, {\it Phys. Rev.}
%{\bf C34}, 1729 (1980).


\begin{thebibliography}{0}
\bibitem{jimsmith}A. Abashian, R, J.
Abrams, D. W. Carpenter, B. M. K. Nefkens and J. H. Smith, {\it Phys. Rev.
Lett.} {\bf 13},243 (1964)

\bibitem{NewPenta}
M. Karliner and H.J. Lipkin,
{\it Phys.\ Lett.}  {\bf B575},  249 (2003).

\bibitem{cryptopen} Marek Karliner and Harry J. Lipkin.
%Why the $\Theta^+$ is seen in some experiments and not in others
% a possible explanation
hep-ph/0405002 and the experimental references therein.

\bibitem{jenmalt}
Byron K. Jennings and Kim Maltman, hep-ph/0308286.

\bibitem{Landsberg:1999wn}
L.~G.~Landsberg, {\it Phys.Rept.}320 223 (1999);
%``The search for pentaquark baryon with hidden strangeness,''
hep-ex/9910048.
%%CITATION = HEP-EX 9910048;%%

\bibitem{Karliner:2004qw}
M.~Karliner and H.~J.~Lipkin,
%``The narrow width of the Theta+: A possible explanation,''
{\it Phys.\ Lett.} {\bf B586}, 303 (2004)
hep-ph/0401072.
%%CITATION = HEP-PH 0401072;%%

\bibitem{Nambu}{Y. Nambu, in Preludes in Theoretical Physics, edited by
A. de Shalit, H. Feshbach and L. Van Hove, (North-Holland Publishing Company,
Amsterdam, 1966), p. 133}
\bibitem{LipkTriEx}{H.J. Lipkin,
{\it Phys. Lett.} {\bf B45},  267 (1973)}
\bibitem{DGG}{A. De Rujula, H. Georgi and S.L. Glashow, {\it Phys. Rev.} {\bf
D12}, 147 (1975)}
\bibitem{Jaffe}{R. L. Jaffe, {\it Phys. Rev. Lett.} {\bf 38},  195 (1977)}
\bibitem{liptet}{H.J. Lipkin,
{\it Phys. Lett.} {\bf B70},  113 (1977)} 
\bibitem{Pent97} {Harry J. Lipkin,
{\it Nucl. Phys.} {\bf A625},  207 (1997)}
\bibitem{E791Col}{E.M. Aitala et al.,
%"Search for the Pentaquark via the $P^0_{{\bar c}s} \rightarrow \phi \pi p$
%Decay" E791 Collaboration,
FERMILAB-Pub-97/118-E, {\it Phys. Lett.}  {\bf B448}, 303  (1996). }

\end{thebibliography}
\end{document}